
%
%
%
%
\documentstyle{amsppt}
\NoBlackBoxes
%
%
\catcode`\@=11
\def\logo@{}
\catcode`\@=13
%


\def\gr#1{{\goth #1}}



	\def\grg{{\gr g}}

	\def\grs{{\gr s}}
	
	\def\gru{{\gr u}}


\def\gam{\gamma}

\def\sig{\sigma}		
\def\vphi{\varphi}
\def\ome{\omega}		\def\Ome{\Omega}
\def\nchi{\hbox{\raise 2.5pt\hbox{$\chi$}}}


\def\CalF{{\Cal F}}

\def\CalM{{\Cal M}}
\def\CalN{{\Cal N}}
\def\CalO{{\Cal O}}
\def\CalP{{\Cal P}}


		\def\bfP{{\bold P}}
		
		\def\bfR{{\bold R}}

		\def\bfZ{{\bold Z}}

\def\atil{\tilde a}
\def\btil{\tilde b}
\def\ctil{\tilde c}

\def\ztil{\tilde z}

\def\Ctil{\widetilde C}

\def\Jtil{\widetilde J}

\def\Ptil{\widetilde P}

\def\Wtil{\widetilde W}


		\def\Fbar{\overline{F}}
		
\def\hbar{\bar h}

\def\xbar{\bar x}


		\def\Jhat{\widehat {\mathstrut J}}

		\def\What{\widehat {\mathstrut W}}


%
%

\def\r1{\sqrt{-1}}
\def\lb{\linebreak}

\def\lbd{\lambda}
\def\mx{\pmatrix}
\def\emx{\endpmatrix}
\def\smx{\left(\smallmatrix}
\def\esmx{\endsmallmatrix\right)}
\def\cx{\text{\bf C}}
\def\rx{\bfR}

\def\tp{^{\text{\bf T}}}
\define\union#1#2{\overset{#2}\to{\underset{#1}\to \cup}}
\def\lo{\Cal L_{0}}
\def\li{\Cal L_{1}}

\def\Ln{\Cal L}

\def\bj{\beta_j}

\def\wt{\wtil}

\def\tr{\text{tr}}
\def\al{\alpha_i}
\def\ai{\al}

\def\aj{\alpha_j}

\def\kmm{\kern -6pt}
\def\ktm{\kern 7pt}
\def\kmc{\kern -7pt}
\def\ktmi{\kern 5pt}
\def\kmmi{\kern -6pt}
\def\kenmi{\kern 3pt}
\def\knenmi{\kern -4pt}

\redefine\cite#1{{\bf[#1]}}

\def \smaller {\eightpoint}
\def\tr{\operatorname{tr}} 
\TagsOnRight
\parindent=8 mm
\magnification \magstep1
\hsize = 6.25 true in
\vsize = 8.5 true in
\hoffset = .2 true in
\parskip=\medskipamount
\topmatter
\title
 Isospectral flow in Loop Algebras \\
and Quasiperiodic solutions of the Sine-Gordon Equation${}^\dag$
\endtitle
\rightheadtext{Isospectral flow for the Sine-Gordon Equation}
\leftheadtext{J. Harnad and M.-A.~Wisse}
\author
J. Harnad${}^1$
and
M.-A. Wisse${}^2$
\endauthor
\endtopmatter
\footnote""{${}^1$Department of Mathematics and
  Statistics, Concordia University, Montr\'eal, P.Q. and \lb
  Centre de Recherches Math\'ematiques,
  Universit\'e de Montr\'eal, C.P. 6128-A,
  Montr\'eal, P.Q. , Canada H3C 3J7}
\footnote"" {${}^2$D\'epartement de math\'ematiques
et de statistique, Universit\'e de Montr\'eal, C.P. 6128-A,\lb
  Montr\'eal, P.Q., Canada H3C 3J7}
\footnote"" {${}^\dag$Research supported in part by the
Natural Sciences Engineering Research Council of Canada and the Fonds FCAR
du Qu\'ebec}
\bigskip
\centerline{\bf Abstract}
\bigskip
\baselineskip=10pt
\centerline{
\vbox{
\hsize= 5.5 truein
{\smaller The sine-Gordon equation is considered in the hamiltonian framework
provided by the Adler-Kostant-Symes theorem. The phase space, a finite
dimensional
coadjoint orbit in the dual space $\grg^*$ of a loop algebra $\grg$, is
parametrized by a finite dimensional symplectic vector space $W$ embedded
into $\grg^*$ by a moment map. Real quasiperiodic solutions
are computed
in terms of theta functions using a Liouville generating function which
generates a canonical transformation to linear coordinates on the Jacobi
variety of a suitable hyperelliptic curve.
}}}
\baselineskip=16pt \bigskip \bigskip

\def\wt{\widetilde}
\def\wh{\widehat}
\def\bj{\beta_j}
\document
\bigskip
\noindent {\bf 1. Introduction}
\smallskip

Quasiperiodic flows and finite-gap solutions of the sine-Gordon equation
$$
\frac{\partial^2u}{\partial x^2}-\frac{\partial^2u}{\partial t^2}
=m^2 \sin(u),\ m\in\rx \tag{1.1}
$$
have been studied by many authors and derived in a variety of ways
\cite{KK, C, M, FM, DN, Da, AA, P, FFS, Sm}. Real solutions for two gaps were
identified in
\cite{DN} and for any number of gaps in \cite{Da} using Baker-Akhiezer
functions. In \cite{FM, AA} the flow was explicitly linearized on  the
Jacobi variety of a hyperelliptic Riemann surface. A Liouville generating
function leads to the linearization in \cite{AA}, where quasiperiodic solutions
to equation \thetag{1.1} are determined in terms of the integrated flows of a
completely integrable finite-dimensional Hamiltonian
system. However, the recovery of real solutions remains incomplete in
\cite{FM} and \cite{AA} and the unifying r\^ole of the loop algebra
$\wt{\grs\gru}(2)$ or the Adler-Kostant-Symes (AKS) theorem \cite{A, K, Sy}
does not appear.

In this paper we shall obtain such solutions using a general approach based
upon moment maps from finite dimensional symplectic vector spaces into
loop algebras as developed in \cite{AHP, AHH1, AHH2}. The AKS theorem and the
Liouville-Arnold
integration method in the context of loop algebras are central to this
approach.

As shown in \cite{AHP}, finite dimensional coadjoint orbits $\CalO$ in loop
algebras
may be parametrized
by symplectic vector spaces via such moment map embeddings.
Equations of the type \thetag{1.1}  are
recovered as compatibility conditions for a pair of Hamiltonian Lax equations
on
$\CalO$, as given by the AKS theorem. The flow on $\CalO$ generates, as usual,
a flow of line bundles defined on the underlying invariant spectral curve.
The divisor coordinates defining these line bundles give a system of
Darboux coordinates on $\CalO$ according to the general scheme developed in
\cite{AHH2}. A Liouville generating function yields the
canonical transformation to linear coordinates on the Jacobi variety of the
spectral curve via the Abel map.

The new element here, relative to the generic case \cite{AHH2}, is the passage
to a
hyperelliptic curve that is not exactly the spectral curve of the flow in
$\CalO$,
but rather a quotient by an involutive automorphism, extended by an additional
two branch points. The quotienting is necessary because the structure of
the Lax pairs implies that the flow is on the twisted loop algebra
$\wh{\grs\gru}(2)^*$, while the additional branch points are necessary
to raise the genus of the curve so as to identify the Jacobi variety with the
Liouville-Arnold torus. It also becomes clear in this approach that real
solutions of \thetag{1.1} are equivalent to the choice of a real submanifold in
$\CalO$ corresponding to a coadjoint orbit of the twisted loop algebra
$\wh{\grs\gru}(2)^+$.

\bigskip
\noindent {\bf 2. Darboux Coordinates for the Sine-Gordon Equation}
\smallskip

First we apply in detail the moment map embedding method developed in
\cite{AHP} to the case of the loop algebra $\wt{\grs\gru}(2)$, in order to
obtain suitable ``Cartesian'' coordinates on rational coadjoint orbits.
Consider the vector space $\cx^{2\times 2}$ consisting of $2 \times 2$ complex
matrices $F$. Let $F_i=(x_i,y_i),\,i=1,2$ be the rows of $F$. A real
symplectic structure on $\cx^{2 \times 2}$ is given by
$$
\wt{\Ome}=dF_2\wedge d\Fbar_1\tp + d\Fbar_2 \wedge dF_1\tp. \tag{2.1}
$$
Now consider the submanifold
$$
\wt{\CalM}=\{F\in\cx^{2 \times 2}\,:\,F_1\neq 0,\,F_2\neq 0,\,
F_2 \Fbar_1\tp=0\}.\tag{2.2}
$$
This is not a symplectic manifold, but on $\wt{\CalM}$ we have an
action of $\cx^*$ given by
$$
h(F)=\mx hF_1 \\ \hbar^{-1} F_2 \emx,\quad h\in \cx^* \tag{2.3}
$$
which preserves the restriction $\wt{\Ome}|_{\wt{\CalM}}$ of the symplectic
form \thetag{2.1} to $\wt{\CalM}$ and whose orbits are the null leaves of
$\wt{\Ome}|_{\wt{\CalM}}$. Thus $\wt{\Ome}$ projects to a symplectic form on
the quotient manifold $\wt{\CalM}/\cx^*$. The submanifold $\Wtil$ of
$\wt{\CalM}$ given by
$$
\Wtil=\{F\in\wt{\CalM}\,\vert\,y_1=\xbar_2,\,y_2=-\xbar_1\}\tag{2.4}
$$
is transversal to the $\cx^*$-orbits, intersecting each in 2 points, giving
a two-sheeted cover of $\wt{\CalM}/\cx^*$, which is identified with the
quotient $\Wtil/{\bfZ}_2$ of
$\Wtil$ by the ${\bfZ}_2$ action $\pm:(x_1,x_2)\mapsto \pm(x_1,x_2)$.
Thus, restricting \thetag{2.1} to $\Wtil$, the manifold $\wt{\CalM}/\cx^*
\equiv \Wtil/{\bfZ}_2$
inherits the symplectic structure
$$
\wt{\ome}|_{\Wtil}= 2(d\gamma\wedge d\bar{\varphi}
+ d\bar{\gamma}\wedge d\varphi).\tag{2.5}
$$
where $(\varphi,\gam)=\pm(x_1,x_2)$, which is well defined
on $\wt{\CalM}/\cx^* \equiv \Wtil/{\bfZ}_2$.
This is just the Marsden-Weinstein reduction at the zero level set of the
moment map $F_2\Fbar_1\tp$ generating the action \thetag{2.3}.

Now, considering $\cx^{(2N)\times 2}$ as the Cartesian product of $N$
copies of $\cx^{2\times 2}$, with coordinates $\smx F_{2i-1} \\ F_{2i}\esmx=
\smx x_{2i-1} & y_{2i-1} \\ x_{2i} & y_{2i} \esmx$ in each $2\times 2$ block,
$\wt{\Ome}$ generalizes to the following symplectic form on
$\cx^{(2N)\times 2}$:
$$
\Ome=\sum_{i=1}^N \left(dF_{2i}\wedge d\Fbar_{2i-1}\tp +
d\Fbar_{2i} \wedge dF_{2i-1}\tp\right).\tag{2.6}
$$
On the submanifold
$$
\CalM=\{F\in\cx^{(2N)\times 2}\,\vert\,F_{2i-1}\neq 0,\,F_{2i}\neq 0,\,
F_{2i}\Fbar_{2i-1}\tp=0,\,i=1,\dots,N\}\tag{2.7}
$$
a symplectic action of $(\cx^*)^N$ is defined by the action \thetag{2.3} of
$\cx^*$ on each $2\times 2$ block $\smx F_{2i-1} \\ F_{2i} \esmx$.
The analogue of $\Wtil$ \thetag{2.4} is given by
$$
W=\{F\in\CalM\,\vert\,y_{2i-1}=\xbar_{2i},\,y_{2i}=-\xbar_{2i-1},
\,i=1,\dots,N\}. \tag{2.8}
$$
Restricting \thetag{2.6} to $W$ determines the
following
symplectic form on the quotient manifold $\CalM/(\cx^*)^N\equiv
W/({\bfZ}_2)^N$:
$$
\ome=2\sum_{i=1}^N (d\gamma_i\wedge d\bar{\varphi}_i
+ d\bar{\gamma}_i\wedge d\varphi_i)
\tag{2.9}
$$
where $(\varphi_i,\gamma_i)=\pm(x_{2i-1},x_{2i}),\,i=1,\dots,N$.

Let $\wt{\grs\gru}(2)$ be the loop algebra consisting of differentiable maps
from a circle
$K$ centered at the origin of the complex $\lbd$--plane into $\grs\gru(2)$
with Lie bracket evaluated, as usual, pointwise.
There is a splitting into the direct sum of subalgebras
$$
\wt{\grs\gru}(2)=\wt{\grs\gru}(2)^{+}\oplus\wt{\grs\gru}(2)^{-},\tag{2.10}
$$
where $\wt{\grs\gru}(2)^{+}$ is the subalgebra of loops in $\grs\gru(2)$ which
extend holomorphically inside $K$, while $\wt{\grs\gru}(2)^{-}$ is the
subalgebra of loops $X(\lbd)$ which extend
holomorphically outside $K$ and normalized by the
condition $X(\infty)=0$. The Lie algebra $\wt{\grs\gru}(2)$
(resp., $\wt{\grs\gru}(2)^-_0=\{X\in\wt{\grs\gru}(2)\,\vert\,X \text{ extends
holomorphically outside }K\text{, with finite limit }X(\infty)\})$ is densely
embedded into $\wt{\grs\gru}(2)^*$
(resp., $\wt{\grs\gru}(2)^{+*}$) via the nondegenerate, ad-invariant bilinear
form on $\wt{\grs\gru}(2)$ given by
$$
<X,Y>:=\oint_K \frac{\tr(X(\lbd)Y(\lbd))}\lbd d\lbd,\quad
X,Y\in\wt{\grs\gru}(2).\tag{2.11}
$$
Henceforth no notational distinction will be made between elements of
$\wt{\grs\gru}(2)^*$ (resp., $\wt{\grs\gru}(2)^{+*}$) and elements of
$\wt{\grs\gru}(2)$ (resp., $\wt{\grs\gru}(2)_0^-$).

Fix $N$ complex numbers $\ai,\,i=1,\dots,N$ in the interior of $K$. We define
an
injective moment map $\Jtil$ on $\CalM/(\cx^*)^N$, which is locally identified
with $W$, by
$$
\Jtil(\varphi_1,\gamma_1,\dots,\varphi_N,\gamma_N)
= \lbd \sum_{i=1}^N
\frac{\mx -\bar{\gamma}_i\varphi_i & \bar{\gamma}_i^2 \\
-\varphi^2_i & \bar{\gamma}_i\varphi_i\emx}{\ai-\lbd} +
\frac{\mx \gamma_i\bar{\varphi}_i & \bar{\varphi}_i^2 \\
-\gamma_i^2 & -\gamma_i\bar{\varphi}_i \emx}{\bar{\alpha}_i-\lbd}
\in \wt{\grs\gru}(2)^{+*}.\tag{2.12}
$$
The fact that this is really a Poisson map with respect to the Lie-Poisson
structure on $\wt{\grs\gru}(2)^*$ and that, viewed as a moment map, it
generates a Hamiltonian action of the corresponding loop group $\wt{SU}(2)^+$
on $W$, follows from the general results developed in \cite{AHP}. It is also
easily verified that the image of this map coincides with a coadjoint orbit
in $\wt{\grs\gru}(2)^{+*}$.
To obtain real solutions of the sine-Gordon equation, however, we must restrict
to a submanifold of $W$ whose image lies in the dual of the ``twisted''
subalgebra $\wh{\grs\gru}(2)$
of $\wt{\grs\gru}(2)$, consisting
of fixed points of the involutive automorphism
$$
\sig(X)(\lbd)=\smx 1 & 0 \\ 0 & -1 \esmx X(-\lbd) \smx 1 & 0 \\ 0 & -1 \esmx,
\quad X\in\wt{\grs\gru}(2). \tag{2.13}
$$
Analogously to \thetag{2.10} we have a splitting
of $\wh{\grs\gru}(2)=\wh{\grs\gru}(2)^+\oplus\wh{\grs\gru}(2)^-$, as well
as embeddings $\wh{\grs\gru}(2)\hookrightarrow\wh{\grs\gru}(2)^*$ (resp.,
$\wh{\grs\gru}(2)^-_0\hookrightarrow\wh{\grs\gru}(2)^{+*}$) via \thetag{2.11},
where $\wh{\grs\gru}(2)^-_0$ is defined analogously to
$\wt{\grs\gru}(2)^-_0$.

In order that condition \thetag{2.13} be satisfied by the image of $\Jtil$,
the $\ai$'s have to come
in pairs of opposite sign. There are two possibilities: either $\ai$ has a
nonzero real part, in which case we need both $\ai$ and $-\ai$,
or $\ai$ is
purely imaginary. We may reorder these constants so
that $\alpha_{i+p}=-\ai,\,i=1,\dots,p$ for the $\ai$'s with a real part, and
$\alpha_{j}=\r1\bj,\,j=2p+1,\dots,N$ for the purely imaginary constants
$\aj$. On $W$ we must then impose the further constraints
(cf. \cite{AHP, {\rm sec. 5}})
$$
\gather
\varphi_{i+p}\bar{\gamma}_{i+p}=\varphi_i\bar{\gamma}_i,\quad
\bar{\gamma}^2_{i+p}=-\bar{\gamma}^2_{i},\quad
\bar{\varphi}^2_{i+p}=-\bar{\varphi}^2_{i},\quad i=1,\dots,p
\tag{2.14a}\\
\varphi_j\bar{\gamma}_j=-\bar{\varphi}_j\gamma_j,\quad
\varphi_j^2=-\gamma_j^2,\quad j=2p+1,\dots,N.\tag{2.14b}
\endgather
$$
Solving these constraints we obtain
$$
\gather
\varphi_{i+p}=\r1 \varphi_i,\quad \gamma_{i+p}=\r1 \gamma_i,\quad
i=1,\dots,p\tag{2.15a} \\
\varphi_j=\r1\gamma_j,\quad j=2p+1,\dots,N. \tag{2.15b}
\endgather
$$
Let $\What$ be the submanifold of $W$ given by constraints \thetag{2.14a.b}
and \thetag{2.15a,b}.
The restriction $\hat\ome$ of $\ome$ to $\What$ is
$$
\hat\ome=
4\sum_{i=1}^p(d\gamma_i\wedge d\bar\varphi_i
+ d\bar\gamma_i \wedge d\varphi_i) +
4\r1 \sum_{j=2p+1}^N d\bar\gamma_j\wedge d\gamma_j, \tag{2.16}
$$
showing that $\hat\ome$ is of maximal rank, and $\What$ a symplectic
subspace.

In terms of the reduced coordinates, the restriction $\Jhat$ of $\Jtil$ to
$\What$ is
$$
\Jhat=2\lbd \mx b(\lbd) & c(\lbd) \\ -\bar{c}(\bar\lbd) & -b(\lbd) \emx,
\tag{2.17}
$$
with $b(\lbd),\,c(\lbd)$ given by
$$
\aligned
b(\lbd)&=\lbd\sum_{i=1}^p \left(\frac{-\varphi_i \bar\gamma_i}{\ai^2-\lbd^2} +
\frac{\bar\varphi_i \gamma_i}{\bar\alpha_i^2-\lbd^2}\right) +
\r1\lbd\sum_{j=2p+1}^N \frac{\left\vert\gamma_j\right\vert^2}{\bj^2+\lbd^2}\\
c(\lbd)&=\sum_{i=1}^p \left(\frac{\ai\bar\gamma_i^2}{\ai^2-\lbd^2}+
\frac{\bar{\alpha}_i\bar\varphi_i^2}{\bar\alpha_i^2-\lbd^2}\right)
-\r1\sum_{j=2p+1}^N\frac{\bj\bar\gamma_j^2 }{\bj^2+\lbd^2}.
\endaligned
\tag{2.18}
$$
Thus, $\Jhat$ takes values in the dual of the twisted loop algebra
$\wh{\grs\gru}(2)^+$.
On $\wh{\grs\gru}(2)^{+*}$, define the ring of functions
$$
\split
\CalF_+=\{\Phi\in C^\infty(\wh{\grs\gru}(2)^{+*})\ \vert\
&\Phi=\wh{\Phi}\vert_{\wh{\grs\gru}(2)^{+*}},\,\wh{\Phi}\in
C^\infty(\wh{\grs\gru}(2)^*)\\&\mu([d\wh{\Phi}(\mu),X])=0,\,\forall
\mu\in\wh{\grs\gru}(2)^{*},\,X\in\wh{\grs\gru}(2)\}.\endsplit\tag{2.19}
$$
That is, $\CalF_+$ consists of the restriction to $\wh{\grs\gru}(2)^{+*}$ of
the infinitesimally $Ad^*$-invariant functions on $\wh{\grs\gru}(2)$.
It follows from the AKS theorem that the Hamiltonians in $\CalF_+$ commute in
the Lie-Poisson
structure of $\wh{\grs\gru}(2)^{+*}$. Choose Hamiltonians $H_\xi,H_\eta\in
\CalF_+$ defined by
$$
\aligned
H_\xi(X)&=\frac 12 \tr\left(\frac{a(\lbd)}{\lbd^2}(X(\lbd)+\lbd Y)^2\right)_0\\
H_\eta(X)&=-\frac 12 \tr\left(\frac{a(\lbd)}{\lbd^{2N}}(X(\lbd)+\lbd Y)^2
\right)_0
\endaligned\tag{2.20}
$$
where $a(\lbd)=\prod_{i=1}^p[(\lbd^2-\ai^2)(\lbd^2-\bar\alpha_i^2)]
\prod_{j=2p+1}^N(\lbd^2+\bj^2)$ and $Y=\smx 0 & -1 \\ 1 & 0 \esmx$.
According to the AKS theorem, Hamilton's equations for
$$
\CalN(\lbd)=\Jhat + \lbd Y \tag{2.21}
$$
are given by
$$
\aligned
\frac d{d\xi}\CalN &= -\left[dH_\xi(\CalN)_-,\CalN\right] \\
\frac d{d\eta}\CalN &= \left[dH_\eta(\CalN)_+,\CalN\right],
\endaligned
\tag{2.22}
$$
where the ``+'' (resp., ``--'') subscript denotes projection to
$\wh{\grs\gru}(2)^+$ (resp., $\wh{\grs\gru}(2)^-$).
Setting
$$
\Ln(\lbd)=\frac{a(\lbd)}\lbd \CalN(\lbd)=\lbd^{2N-1}\lo+\lbd^{2N-2}\li+\dots+
\lbd^0\Ln_{2N-1}+a(\lbd)Y,\tag{2.23}
$$
we have
$$
\aligned
dH_\xi(\CalN)_-&=\frac 1\lbd (\Ln_{2N-1} + a(0)Y) \\
dH_\eta(\CalN)_+&= \lo + \lbd Y,
\endaligned\tag{2.24}
$$
and the Lax equation for $\Ln(\lbd)$ is obtained by multiplication of
\thetag{2.22} by $\frac{a(\lbd)}\lbd$.

The flow leaves invariant the spectral curve
with affine part given by
$$
\CalP(\lbd,z)=\det(\Ln(\lbd)-zI)=0. \tag{2.25}
$$
Expanding \thetag{2.25} and using the fact that the rank of the residues of
$\Jhat$ is equal to 1, we may write
$$
\gather
\CalP(\lbd,z)=z^2+a(\lbd)P(\lbd)=0 \tag{2.26a} \\
P(\lbd)=P_0+\lbd^2 P_1 + \dots + \lbd^{2N-2} P_{N-1} +\lbd^{2N}. \tag{2.26b}
\endgather
$$
The coefficients $P_i$ are all in $\CalF_+$; in particular,
$$
H_\xi=-P_0,\qquad H_\eta=P_{N-1}. \tag{2.27}
$$
The $P_i$ are thus constant along the flows of all Hamiltonians
in $\CalF_+$. We may choose a level set
$$
P_0=\det(\Ln_{2N-1}+a(0)Y)\equiv\frac{m^2}{16} \in \rx, \tag{2.28}
$$
which implies that we may write
$$
\Ln_{2N-1}+a(0)Y=\frac m4 \mx 0 & e^{\r1 u} \\ -e^{-\r1 u} & 0 \emx, \tag{2.29}
$$
where $u$ is real. From \thetag{2.29} we see that $u$ may be written in terms
of the
coordinates $(\varphi_i,\gamma_i,i=1,\dots,p;\,\gamma_j,j=2p+1,\dots,N)$
as follows:
$$
e^{\r1 u}=a(0)(c(0)-1). \tag{2.30}
$$
{}From the compatibility conditions for equations \thetag{2.22} it now follows
that $u$
satisfies equation \thetag{1.1}, where $\xi=x+t$ and $\eta=x-t$.

\noindent[{\it Remark.} Equation \thetag{1.1} may equivalently be viewed as the
compatibility conditions for the $x$ and $t$ flows determined by the two
Hamiltonians $H_x:=-H_\xi-H_\eta,\,H_t:=H_\xi-H_\eta$.]

\bigskip
\noindent{\bf 3. Quasiperiodic Solutions for the Sine-Gordon equation}
\smallskip
The hyperelliptic spectral curve \thetag{2.25}  has genus $g=2N-1$.
However,
it is invariant under the involution $(z,\lbd)\mapsto(z,-\lbd)$, and hence is a
two-sheeted covering of the hyperelliptic curve $C'$, with genus $N-1$, whose
affine  part given is by
$$
z^2+\atil(E)\Ptil(E)=0, \tag{3.1}
$$
where $\Ptil(\lbd^2)=P(\lbd)$ and $\atil(\lbd^2)=a(\lbd),\,\lbd^2=E$.
We also define functions $\btil,\ctil$ by
$\btil(E)=b(\sqrt E),\,\ctil(E)=c(\sqrt E)$.
In order to apply the Jacobi inversion method we shall need
a hyperelliptic curve $\Ctil$ of genus $N$ with affine part given
by
$$
\ztil^2+E\atil(E)\Ptil(E)=0.\tag{3.2}
$$
This is obtained from $C'$ by setting $\ztil=z\lbd$, which adds branch points
at $E=0$ and $E=\infty$. Following the general
method for the introduction of ``spectral Darboux coordinates'' developed in
\cite{AHH2}, we define on $\Ctil$ a divisor of degree $N$ with coordinates
$(E_\mu,\zeta_\mu)_{\mu=1,\dots,N}$. The $E_\mu$
are given by the equation
$$
\ctil(E_\mu)-1=0,\tag{3.3}
$$
or, equivalently
$$
\sum_{i=1}^p \left(\frac{\ai\bar\gamma_i^2}{\ai^2-E}+
\frac{\bar{\alpha}_i\bar\varphi_i^2}{\bar\alpha_i^2-E}\right)
-\r1\sum_{j=2p+1}^N\frac{\bj\bar\gamma_j^2 }{\bj^2+E}-1=
-\frac{\prod_{\mu=1}^{N}(E-E_\mu)}{\atil(E)}.\tag{3.4}
$$
These may be viewed as complex hyperelliptic coordinates if the
$(\varphi_i,\gamma_i,i=1,\dots,p;\allowmathbreak
\gamma_j,j=2p+1,\dots,N)$ are interpreted as Cartesian
coordinates on the submanifold of $\CalM/(\cx^*)^N$ determined by
the constraints \thetag{2.14a,b}.
The canonically conjugate coordinates $\zeta_\mu$ are defined by
$$
\zeta_\mu=\sqrt{-\frac{\Ptil(E_\mu)}{E_\mu \atil(E_\mu)}}=
-\frac{2\btil(E_\mu)}{\sqrt{E_\mu}},\tag{3.5}
$$
i.e. by the eigenvalues of the matrix $\frac{\CalN(\lbd)}{\lbd^2}$ at
$\lbd^2=E_\mu$.
Comparing with \thetag{2.30} we see that
$$
\gather
e^{\r1 u}=-\prod_{\mu=1}^{N}(-E_\mu)\tag{3.6}\\
u=-\r1\left(\sum_{\mu=1}^N \ln(-E_\mu)-\pi\right).\tag{3.7}
\endgather
$$

\proclaim{\bf Proposition 3.1}
The coordinates $(E_\mu,\zeta_\mu)_{\mu=1,\dots,N}$ form a Darboux coordinate
system on the coadjoint orbit passing through $\CalN(\lbd)$. The corresponding
symplectic form is
$$
\ome_\CalN=\sum_{\mu=1}^{N} dE_\mu\wedge d\zeta_\mu=-d\theta. \tag{3.8}
$$
\endproclaim
\demo{Proof}
Computing the differentials of the residues of \thetag{3.4} and summing up
we find, using \thetag{3.5},
$$
4\sum_{i=1}^p(\varphi_i d\bar\gamma_i-\gamma_i
d\bar\varphi_i)+4\sum_{j=2p+1}^N\gamma_j d\bar\gamma_j=\sum_{\mu=1}^{N}
\zeta_\mu dE_\mu
= \theta.\qed\tag{3.9}
$$
\enddemo

\noindent[{\it Remark:} This result could also have been obtained as in
\cite{AHH2}, by computing implicitly the Lie-Poisson brackets of
$(E_\mu,\zeta_\mu)$ on the coadjoint orbit passing through $\CalN(\lbd)$.]

The restriction of \thetag{3.8} to the invariant level sets $P_i=c_i=$const.,
$i=0,\dots,N-1$ is identically zero; i.e. this defines a Lagrangian
submanifold. Hence in a neighborhood of $P_i=c_i$,
the one-form $\theta=dS$ may be integrated on the leaves of the Lagrangian
foliation
as usual to yield the Liouville generating function
$$
S(P_i,E_\mu)=\sum_{\mu=1}^{N} \int_{E_0}^{E_\mu}
\sqrt{-\frac{\Ptil(E)}{E\atil(E)}}dE,
\tag{3.10}
$$
where $E_0\in \Ctil$ is a suitably chosen base point.
Derivation of $S$ with respect to the $P_i$'s gives the conjugate
coordinates $Q_i$, in terms of which the flows of the
Hamiltonians in the ring generated by the $P_i$'s are linear.
$$
Q_i=\frac{\partial S}{\partial P_i}= -\frac 12
\sum_{\mu=1}^{N} \int_{E_0}^{E_\mu}
\frac{E^i}{\sqrt{-E\atil(E)\Ptil(E)}}dE.\tag{3.11}
$$
Hamilton's equations for $H_\xi,H_\eta$ are thus integrated to give
$$
\sum_{\mu=1}^{N} \int_{E_0}^{E_\mu} \frac{E^i}{\sqrt{-E\atil(E)\Ptil(E)}}dE
=C_i+2\delta_{i,0}\xi-2\delta_{i,N-1}\eta, \tag{3.12}
$$
on the Jacobi variety of $\Ctil$.

If the
holomorphic differentials
$$
\ome_i=\frac{E^i}{\sqrt{-E\atil(E)\Ptil(E)}}dE\tag{3.13}
$$
were normalized, the left hand side of \thetag{3.12} would just be the Abel
map.
Since the  $\ome_i$ form a  basis of holomorphic
differentials for $\Ctil$, choosing a basis $(a_i,b_i)_{i=1,\dots,N}$ of
$H_1(\Ctil,\text{\bf Z})$ such that $a_i\cdot a_i=b_i\cdot b_i=0,\,
a_i\cdot b_j=\delta_{ij}$, the matrix $M$ of integrals over the $a$-cycles of
the differentals $\ome_i$,
$$
M_{ij}=\oint_{a_i}\ome_j,\tag{3.14}
$$
is invertible.
Multiplying equation \thetag{3.12} by $M^{-1}$, the flow for equation
\thetag{1.1} is linearized on the Jacobi-variety of the curve $\Ctil$ via
the
Abel map
$$
A(p_1+\dots+p_N)=U\eta+V\xi+B,\tag{3.12'}
$$
where $p_\mu \in\Ctil$ are the points with coordinates $(\zeta_\mu,
E_\mu)$, and $U,V,B$ are constant vectors in $\cx^N$ obtained by applying
$M^{-1}$ to
the vectors with components $-2\delta_{i,N-1}$, $2\delta_{i,0}$ and $C_i$,
respectively, appearing in equation \thetag{3.12}. It remains to explicitly
compute
$u(\xi,\eta)$ as given by equation \thetag{3.7} from this linear flow.
Let $\Theta$ be the theta function corresponding to the hyperelliptic curve
$\Ctil$ and $\kappa$, the Riemann constant. From \thetag{3.7} follows, in
a standard way (cf. \cite{GH, Du, AHH2 {\rm (Cor. 1.7)}}):

\proclaim{Proposition 3.2}
$$
u=-2\r1\ln\frac{\Theta(A(0)-U\eta-V\xi-B-\kappa)}
{\Theta(A(\infty)-U\eta-V\xi-B-\kappa)}
+C \tag{3.15}
$$
where $C$ is an integration constant independant of $\eta$ and $\xi$.
\endproclaim

\demo{Proof}
As usual we consider $\Ctil$ as a two sheeted branched cover of $\bfP^1(\cx)$
with
projection mapping $\pi:\Ctil\rightarrow\bfP^1(\cx)$ and branch locus $\Cal B$.
Let $\{a_i,b_i\},\,i=1,\dots,N$  be a basis of
$H_1(\Ctil,\bfZ)$ with common base point $p_0$. Let $\pi(a_i)$ and $\pi(b_i)$
be
the projections to  $\bfP^1(\cx)$. We suppose that the $a_i$'s are chosen such
that the winding numbers $n(\pi(a_i),0)$ and $n(\pi(a_i),\infty)$ are zero
and such
that their intersection indices are given by $a_i\cdot a_j=b_i\cdot
b_j=0,\,a_i\cdot
b_j=\delta_{ij}$.
Consider a polygonization $\Delta$ of $\Ctil$ with respect to
$(a_i,b_i)$. Assume that $F(E)=\Theta(A(E)-U\eta-V\xi-B-\kappa)$ is not
identically
zero on $\Ctil$ (which holds for generically chosen $B$) and that
$E_\mu\neq0,\infty,\,\mu=1,\dots,N$. On $\Delta$ choose
integration paths $c_0$ (resp., $c_\infty$) from one of the representatives
of $p_0$ on
$\Delta$ (i.e. one of the vertices of $\Delta$) to $0$ (resp., $\infty$).
Cut along these paths to obtain a
polygonization $\wt{\Delta}$. On $\wt{\Delta}$ the differential
$$
\varphi=\ln(-E)d\ln F(E)\tag{3.16}
$$
is well defined and meromorphic. It is easily computed that
$$
\sum_{\mu=1}^{N}\ln(-E_\mu)=\oint_{\partial \wt{\Delta}} \varphi,\tag{3.17}
$$
whereas the right hand side of \thetag{3.17} yields (see e.g. \cite{GH})
$$
\aligned
\int_{a_i+a_i^{-1}}\vphi&=\text{constant in }\eta\text{ and }\xi\\
\int_{b_i+b_i^{-1}}\vphi&=0\\
\int_{c_0+c_0^{-1}}\vphi+\int_{c_\infty+c_\infty^{-1}}\vphi&=
2\pi\r1\ln\frac{\Theta(A(0)-U\eta-V\xi-B-\kappa)}
{\Theta(A(\infty)-U\eta-V\xi-B-\kappa)}.
\endaligned\tag{3.18}
$$
Summing up these terms and substituting in \thetag{3.7}
yields the desired result.\qed
\enddemo

\bigskip
\noindent{\bf Conclusions:}
\smallskip
The form of the quasiperiodic solutions \thetag{3.15} agrees with that obtained
by other authors (e.g., \cite{FM, DN}) who have studied the sine-Gordon
equation
by a variety of methods. The new element presented here is the placing of these
solutions entirely within the framework of isospectral flows  in loop algebras
and the AKS theorem. The reality conditions and reductions related to  the
invariance of the spectral curve under involutions follow naturally in this
approach from the use of the twisted loop algebra $\wh{\grs\gru}(2)$. The
Darboux coordinates leading to a completely separated Liouville generating
function $S$ given by \thetag{3.10} and the linearization of the flow via the
Abel map
are seen as illustrative cases of the ``spectral Darboux coordinate''
method developed in \cite{AHH2}.

\bigskip
\noindent{\it Acknowledgements}: The authors are pleased
to acknowledge helpful discussions with J.~Tafel
relating to the contents of this paper.

\bigskip
\centerline{\smc References}

\medskip
\item{\bf [A]} Adler, M., On a Trace Functional for Formal
Pseudo-Differential Operators and the Symplectic Structure of the Korteweg-
de Vries Equation,  {\sl Invent\. Math\.} {\bf 50} (1979), 219-248.
\item{\bf [AA]} Al'ber, S.J., Al'ber, M.S.,
Hamiltonian Formalism for Nonlinear Schr\"odinger Equations and
Sine-Gordon Equations, {\sl J\. London Math\. Soc\. (2)} {\bf 36} (1987),
176--192.
\item{\bf [AHH1]} Adams, M.R., Harnad, J. and Hurtubise, J.,
Isospectral Hamiltonian Flows in Finite and Infinite Dimensions II.
Integration of Flows,
{\sl Commun\. Math\. Phys\.} {\bf 134} (1990), 555--585.
\item{\bf [AHH2]} Adams, M.R., Harnad, J. and Hurtubise, J.,
Darboux Coordinates and Liouville-Arnold Integration in Loop Algebras,
{\sl CRM-preprint} (1991);
Coadjoint Orbits, Spectral Curves and Darboux Coordinates,
in: ``The Geometry of Hamiltonian Systems'', ed\. T. Ratiu,
Publ. MSRI Springer-Verlag, New York (1991);
Liouville Generating Function for Isospectral Hamiltonian
Flow in Loop Algebras, in:  ``Integrable and
Superintergrable Systems'', ed\. B. Kuperschmidt, World Scientific,
Singapore (1990)
\item{\bf [AHP]} Adams, M.R., Harnad, J. and Previato, E.,
Isospectral Hamiltonian Flows in Finite and Infinite Dimensions I.
Generalized Moser Systems and Moment Maps into Loop Algebras,
{\sl Commun\. Math\. Phys\.} {\bf 117} (1988), 451--500.
\item{\bf [C]} Cherednik, I.V.,
Reality Conditions in Finite Zone Integration, {\sl Dokl\. Akad\. Nauk SSR}
{\bf 25}, no.~5 (1980), 1104--1108.
\item{\bf [Da]} Date, E.,
Multi-Soliton Solutions and Quasi-Periodic Solutions of Nonlinear
Equations of Sine-Gordon Type, {\sl Osaka J\. Math\.} {\bf 19} (1982),
125--158.
\item{\bf [Du]} Dubrovin, B.A., Theta Functions and Nonlinear Equations,
{\sl Russ\. Math\. Surv\.} {\bf 36} (1981), 11--92.
\item{\bf [DN]} Dubrovin, B.A., Natanzon, S.M.,
Real Two Zone Solutions of the
Sine-Gordon Equation, {\sl Funkt\. Anal\. Appl\.} {\bf 16} (1982), 21--33.
\item{\bf [FFS]} Flesh, R., Forest, M.G., Sinha, A.,
Numerical inverse spectral transform for the periodic
sine-Gordon equation: Theta-functions solution and their
linearized stability. {\sl Physica D} {\bf 48} (1991),
169--231. 
D.W.,    Spectral Theory for the periodic
Sine-Gordon Equation: A Concrete Viewpoint,
{\sl J\. Math\. Phys\.} {\bf 23} (1982), no.~7, 1248--1277.
\item{\bf [GH]} Griffith, P. and Harris, J., ``Principles of Algebraic
Geometry'', John Wiley, New York, 1978.
\item{\bf [K]} Kostant, B., The Solution to a Generalized Toda Lattice and
Representation Theory, {Adv\. Math\.} {\bf 34} (1979), 195-338.
\item{\bf [KK]} Kozel, V.A., Kotlyarov, V.P.,
Almost Periodical Solutions of the Equation $u_{tt}-u_{xx}+\sin(u)=0$,
{\sl Dokl\. Akad\. Nauk UkSSR, Ser\. A} {\bf 10} (1978), 878--881 (in
Ukrainian).
\item{\bf [M]} McKean, H.P.,
The Sine-Gordon and Sinh-Gordon Equations on the circle,
{\sl Comm\. Pure Appl\. Math\.} {\bf 34} (1981), 197--257.
\item{\bf [P]} Previato, E. A particle-system model of the sine-Gordon
hierarchy,
Solitons and coherent structures,
{\sl Physica D} {\bf 18}, no.~1-3 (1986), 312--314.
\item{\bf [Sm]} Smirnov, A.O., Real elliptic solutions of the sine-Gordon
equation.
{\sl Math of the USSR Sbornik} {\bf 70} (1991), 231--240.
\item{\bf [Sy]} Symes, W., Systems of Toda Type, Inverse Spectral Problems
and Representation Theory, {\sl Invent\. Math\.} {\bf 59} (1980), 13-51.

\enddocument